\documentclass[10pt,a4paper]{article}

\usepackage[top=1.3in, bottom=1.3in, left=1.5in, right=1.5in]{geometry}
\usepackage[breaklinks]{hyperref}
\usepackage[T1]{fontenc} 
\usepackage[latin1]{inputenc}

\usepackage{aeguill}
\usepackage{amsfonts}
\usepackage{amsmath}
\usepackage{amssymb}
\usepackage{amssymb,stmaryrd}
\usepackage{garamond}
\usepackage[sort&compress]{natbib}
\usepackage[]{url,hyperref}
\usepackage[urw-garamond]{mathdesign}

\usepackage{verbatim} 
\usepackage{graphicx}
\usepackage{tikz}
\usepackage{subfigure}
\usetikzlibrary{arrows} 
\usetikzlibrary{decorations.pathreplacing}
\usetikzlibrary{fit}
\usetikzlibrary{backgrounds,automata}

\title{A Methodology to Engineer and Validate Dynamic Multi-level Multi-agent Based Simulations}

\author{Jean-Baptiste Soyez$^{1,2}$\\Gildas Morvan$^{1,3}$\\Daniel Dupont$^{1,4}$\\Rochdi Merzouki$^{1,2}$\\~\\
$^{1}$Univ. Lille Nord de France\\~\\
$^{2}$ Polytech Lille, LAGIS F-59655 Villeneuve d'Ascq, France\\\href{mailto:jean-baptiste.soyez@polytech-lille.fr}{jean-baptiste.soyez@polytech-lille.fr}\\\href{mailto:rochdi.merzouki@polytech-lille.fr}{rochdi.merzouki@polytech-lille.fr}\\~\\
$^{3}$ UArtois, LGI2A, F-62400, Béthune, France\\ \href{mailto:gildas.morvan@univ-artois.fr}{gildas.morvan@univ-artois.fr}\\~\\
$^{4}$ HEI, UC Lille, F-59046 Lille, France\\ \href{mailto:daniel.dupont@hei.fr}{daniel.dupont@hei.fr}\\~\\
}

\date{}

\begin{document}

\garamond

\maketitle

\begin{abstract}
This article proposes a methodology to model and simulate complex systems, based on IRM4MLS,
a generic agent-based meta-model able to deal with multi-level systems.
This methodology permits the engineering of dynamic multi-level agent-based models,
to represent complex systems over several scales and domains of interest.
Its goal is to simulate a phenomenon using dynamically the lightest representation
to save computer resources without loss of information.
This methodology is based on two mechanisms: (1) the activation or deactivation
of agents representing different domain parts of the same phenomenon and
(2) the aggregation or disaggregation of agents representing the same phenomenon at different scales.
\end{abstract}

\section{Introduction}

Today, more and more engineering  projects try to cope with complex systems.
Complexity can come from the number of represented entities,
their structure, or the fact that information is coming from difference sources and is incomplete.

Agent-based modeling is a very powerful and intuitive framework to study such systems.
However, the limitations of this approach lead to the development of multi-level agent-based modeling (ML-ABM).
It is defined by~\citet[p. 1]{Morvan:2012c} as:
``\textit{Integrating heterogenous ABMs, representing complementary points of view,
so called levels (of organization, observation, analysis, granularity, ... ), of the same system.
Integration means, of course, these ABMs interact but also they can share entities such as environments and agents}''.
From an engineering point of view, ML-ABM reduces the complexity of the problem, so it becomes easier to implement.

In complex systems simulations, it is generally necessary to find a compromise between
the quality of simulations (amount of information or realism) and their resource consumption (used CPU and memory).

A way to deal with this compromise is to use different models, more or less detailed or treating different
aspects of the same phenomenon  and that are (dis)activated  at run-time, according to the context. This article proposes a methodology to engineer
and validate such simulations, based on IRM4MLS, a ML-ABM meta-model proposed by~\citet{Morvan:2011,Morvan:2012b}.

The next section presents recent works in the domain of multi-resolution or multi-level modeling.
Section 3 introduces a generic agent-based meta-model IRM4MLS.
Then, section 4 shows some possibilities offered by IR4MLS to model complex systems in which different domains interact.
Section 5 explains how to construct models with dynamic change of level of detail (LOD), i.e., switching scales or domains of interest.
Section 6 gives a tool to measure the quality of multi-level models endowed with dynamic changes of resolution.
Finally, we expose the conclusions and perspectives of our work in section 7.

\section{Related Works}

In this section, multi-modeling approaches, dealing with models at different scales in an engineering context, are presented.

Multi-Resolution modeling~\citep{Davis:1993} is the joint execution of different models of the same phenomenon within the same simulation
or across several heterogeneous systems. It can inspire our approach if different models can be considered as different levels.
Consistency represents the amount of essential information lost when crossing different models and it is an adapted tool to test
the quality of this approach.

The High Level Architecture~\citep{HLA:2000} (HLA) is a general purpose architecture for distributed
computer simulation systems. Using HLA, computer simulations can interact (communicate data
and synchronize actions) with other computer simulations regardless of the computing platforms.
The interaction between simulations is managed by a Run-Time Infrastructure (RTI).~\citet{Scerri:2010} developed HLA-Repast, a unified agent-based simulation framework, in which concurrent modules with their own temporality can use global variables through centralized services.

Holonic multi-agent systems (HMAS) can be viewed as a specific case of multi-level multi-agent
systems (MAS). The most obvious aspect being the hierarchical organization of levels. However, from a
methodological perspective, differences remain. Most of holonic meta-models focus on organizational and
methodological aspects while ML-ABM is process-oriented. HMAS meta-models have been proposed in various
domains, e.g., ASPECTS~\citep{Gaud:2008} or PROSA\citep{Van-Brussel:1998}. Even if ML-ABM and HMAS
structures are close, the latter is too constrained for the target application of this work.

\citet{Navarro:2011} present a framework to dynamically change the level of detail in agent-based simulation.
That is to say, represent only what is needed during simulation, to save CPU resources and keep the consistency
of the simulation. But this framework is limited because levels form a merged hierarchy, without the possibility of having
two levels at the same scale and communication between levels is not explicitly defined.\\

The possibility for agents to exist in several levels simultaneously is a way to make simulations benefit of a higher power of representation.
It permits to 1) simulate nested entities, 2) create agents with concurrent psychological trends and 3) model complex systems implying various domains.

It is possible to model the coexistence of nested entities at different scales.
Agents present in different levels can be seen as ``gate'' between these levels.
For example, \citet{Picault:2011}, give the example of cell membrane elements that are the ``gates'' between the inside and
 the outside of the cell, i.e., between two scales and exposed to the influences of two different environments.

An agent existing at different levels simultaneously can fulfill a global objective while following its own goals.
In~\citet{Stratulat:2009}, authors decompose, with the MASQ model, agents into two bodies: a physical one (individual) and a social one (collective)
to do this.

Levels can have different temporal dynamics, independently of other levels.
It allows to optimize the execution of complex agents by (dis)activating their bodies at run-time to use the lightest representation~\citep{Soyez:2011}.

Readers interested in a more comprehensive presentation of ML-ABM should refer to~\citet{Gil-Quijano:2012,Morvan:2012c}. 

\section{IRM4MLS}

IRM4MLS is a ML-ABM meta-model proposed by~\citet{Morvan:2011,Morvan:2012b}.
It relies on the influence/reaction model~\citep{Ferber:1996} and its extension to temporal systems, IRM4S~\citep{Michel:2007}.
An interesting aspect of IRM4MLS is that any valid instance can be simulated by a generic algorithm.
The main aspects of this meta-model are presented in this section.

A IRM4MLS model is characterized by a set of levels, $L$, and relations between levels.
Two types of relations are considered: \emph{influence} (agents in a level $l$ are able to produce influences in a level $l'\neq l$) and \emph{perception}
(agents in a level $l$ are able to perceive the state of a level $l'\neq l$).
These relations are respectively formalized by two digraphs, $\langle L, E_I \rangle$ and $\langle L, E_P \rangle$ where $E_I$ and $E_P$ are sets of edges,
i.e., ordered pairs of elements of $L$.
The dynamic set of agents at time $t$ is denoted $A(t)$. $\forall l \in L$, the set of agents in $l$ at $t$ is $A_l(t)\subseteq A(t)$.
An agent acts in a level if a subset of its external state belongs the state of this level.
An agent can act in multiple levels at the same time.
Environment is also a top-class abstraction.
It can be viewed as a tropistic agent with no internal state that produces ``natural'' influences in the level (Fig. 2).

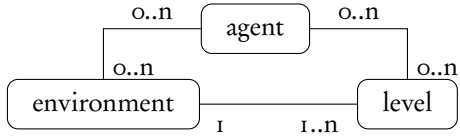
\begin{figure}[t]
\begin{center}

\begin{tikzpicture}

\node(a)[draw] at (0,0) [rounded corners]{
            \begin{tabular}{c}
	environment
\end{tabular}};

\node(b)[draw] at (2,1)[rounded corners]{
            \begin{tabular}{c}
	agent
\end{tabular}};

\node(c)[draw] at (4,0)[rounded corners]{
            \begin{tabular}{c}
	level
\end{tabular}};

\draw (a.north) |- (b.west) 
	node[near end, above] {0..n}
	node[very near start, right]{0..n};
\draw (b.east) -| (c.north)
	node[near start, above] {0..n}
	node[very near end, right]{0..n};
\draw (a.east) -- (c.west)
	node[very near start, below]{1}
	node[near end, below]{1..n};

\end{tikzpicture}
\end{center}
\caption{Central Concepts of IRM4MLS (cardinalities are specifed the UML way)}
\end{figure}

The scheduling of each level is independent: models with different temporalities can be simulated without temporal bias.
On an other hand, only the relevant processes are permitted to execute during a time-step.
A major application of IRM4MLS is to allow microscopic agents (members) to aggregate and form-up lower granularity agents (organizations).
It can be useful to create multiple levels at the same scale to represent different domain parts of the same phenomenon.
In the following, we consider that two levels are at the same scale if they have the same spatial and temporal extents.

\section{Multi-level, Single Scale Simulation}

In this section we give a framework to improve the integration of agents located in different levels
(not necessary at different scales) simultaneously.
Then, we show how to take advantage of this concept to simulate complex systems
while optimizing the use of computer resources.

\subsection{``One Mind, Several Bodies''}

In our approach, inspired by~\citet{Picault:2011}, agents can be present in several levels at the same time.
We propose to decompose agents in a ``central'' \textit{unsituated} part  and a set of $n$ ``peripheral'' parts, each situated in a given level.
Thus, we call \emph{spiritAgent} the unsituated part of the agent which contains its internal state, its decision processes and
that cannot act in a level.
\emph{BodyAgents} in levels $l \in L$ are the situated part of the agent which contains its external state
and the possible actions in its level, like perception of the environment.\\

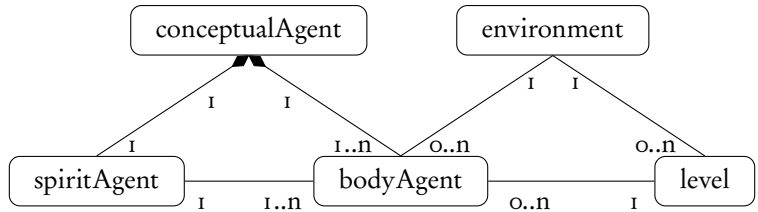
\begin{figure}[t]
\begin{center}

\begin{tikzpicture}

\node(a)[draw] at (0,0) [rounded corners]{
            \begin{tabular}{c}
	spiritAgent
\end{tabular}};

\node(b)[draw] at (2,2)[rounded corners]{
            \begin{tabular}{c}
	conceptualAgent
\end{tabular}};

\node(c)[draw] at (4,0)[rounded corners]{
            \begin{tabular}{c}
	bodyAgent
\end{tabular}};

\node(d)[draw] at (8,0)[rounded corners]{
            \begin{tabular}{c}
	level
\end{tabular}};

\node(e)[draw] at (6,2)[rounded corners]{
            \begin{tabular}{c}
	environment
\end{tabular}};

\draw (e.south) -- (d.north) 
	node[very near end, left] {0..n}
	node[near start, left]{1};

\draw (c.north) -- (e.south)
	node[very near start, right] {0..n}
	node[near end, right]{1};

\draw (c.east) -- (d.west)
	node[near start, below] {0..n}
	node[very near end, below]{1};

\draw (a.east) -- (c.west)
	node[very near start, below]{1}
	node[near end,below]{1..n};

\draw [-diamond] (c.north) -- (b.south)
	node[very near start, left]{1..n}
	node[near end, below]{1};

\draw [-diamond] (a.north) -- (b.south)
	node[very near start, right]{1}
	node[near end, below]{1};

\end{tikzpicture}
\end{center}
\caption{Class diagram of central Concepts of IRM4MLS with separation of situated-or-not agent parts}
\end{figure}

{\bfseries ConceptualAgents} stand for common agents in classical simulation.
{\bfseries SpiritAgents} only contain the \emph{internal state} of the agent and its \emph{decision module}.
{\bfseries BodyAgents} have to be situated in one and only one level. They contain the \emph{external state} of
the agent specific to a level, and an \emph{action module} that indicates: 1) what are
the available actions at a given time and 2) what are their results in term of produced influences.
The \emph{perception process} must be in this action module.
{\bfseries Levels} contain inactive objects that support agent actions.
The only use of {\bfseries Environments} is to produce the \emph{natural influences} of the level (like the gravity force in a physical level).

To obtain valid simulations with such models, a spiritAgent has to be able to access the external state of its conceptualAgent contained in its bodyAgent when it is active
(during the execution of its level). Thus, we can consider the several steps of the life cycle of agents. Each time a bodyAgent is active, 1) it perceives its level
(and others perceptible from this one), 2) it sends a part of these perceptions and the possible actions to the spiritAgent, 3) the spiritAgent modifies
its internal state and 4) indicates the most appropriate action to be accomplished by the bodyAgent, 5) the bodyAgent accomplishes this action which produces influences in
direction of its levels and others possibly influenced by this one.\\

\subsection{Level Temporality}

In this section we explain the possibility to attribute a different temporality to each level and how to adapt it to our models.
IRM4MLS uses the framework of timed event systems~\citep{Zeigler:2000}.
The scheduling is distributed between levels with no constraint on the scheduling mode (step wise or discrete events).
This approach seems more adapted to our problems than the agent one~\citep{Weyns:2003} or the system one~\citep{Michel:2003}.

Our goal to give to agents the longest possible life cycle which stay coherent with the rest of the simulation.
This is done to minimize the computer resources allocated to the agents updating process.
\citet{Morvan:2011} propose an algorithm adapted to IRM4MLS which manage the coupling
between levels with different temporal dynamics. This is made to apply easily the proposed methods above.\\

\begin{figure}[t]
\begin{center}
\includegraphics[width=0.9\textwidth]{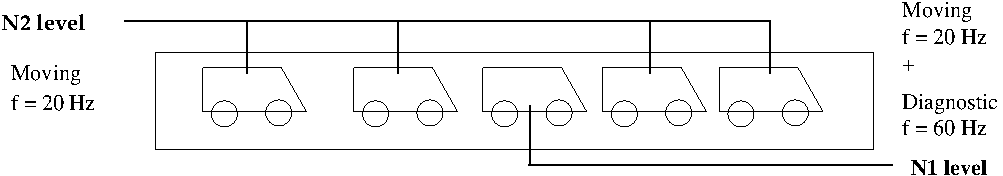}
\caption{Example of Multi-Level ML-ABM with different temporalities}
\end{center}
\end{figure}

The Figure 3 illustrates different constraints which fix the life cycle of agents in a same level.
The frequency of a level is expressed in Hertz, indicating how many times a second, it is necessary to execute the updating process of the dynamic state of a level.
Let imagine that all functions of an agent possess a minimal frequency beyond which their simulation is not
realistic anymore. If a level permits to its agents to
dispose of functions with different frequencies, it adopts the higher one, to keep a correct simulation of the functions with this frequency.
Therefore, in the example of Fig. 3, the frequency of the level $N_1$ is equal to $60 Hz$ because the diagnostic function of the modeled vehicles needs
this minimal frequency.

The other constraint comes from the interactions between levels. If we continue with the previous example, let say that $N_2$ level needs a minimal
frequency equal to $20 Hz$, this frequency could be allocated to $N_2$. However if the $N_1$ level is influenced by $N_2$ and has to calculate the reaction
induced by these influences at a frequency higher than $20 Hz$ (logically less or equal to 60 Hz), it can be necessary to allocate a higher frequency to $N_2$.
Thus, it is necessary to dynamically modify the frequency of a level $N$ and adapt it to the changing needs of the simulation and return it back to its minimal
frequency, defined during the implementation phase.

\section{Dynamic Change of Level of Detail (LOD)}

In this section we give a methodology to apply dynamic changes of LOD in a simulation. First we present the \emph{hierarchical level graph},
which indicates the links between levels and the dis/aggregation functions attached to change the LOD
of simulated entities.
Finally, we specify when and in which conditions dis/aggregation functions can be applied.
In the next part, we give a method to test the quality of the dis/aggregation mechanisms exposed here by measuring the whole consistency
of simulations.

\subsection{Hierachical Level Graph}

Relations between levels are respectively formalized by a digraph,
$\langle L, E_H \rangle$ where $E_H$ are sets of edges, i.e., ordered pairs of elements of $L$.
This digraph whose vertices are levels, is called the \emph{hierarchical level graph}.
This graph indicates how levels are nested and which couple of levels
treats different domain of interest of the same phenomenon.

A \emph{simple edge} represents an \emph{inclusion link} between two levels. For example, an $(l_1,l_2)$ edge signifies that $l_2$ has
higher spatial or temporal extents
than $l_1$. Then the bodyAgents situated in $l_1$ can be aggregated and the resulting aggregate can be instantiated in $l_2$. We note that $l_1\prec l_2$.

A \emph{pair of symmetric edges} means there is a \emph{complementarity link} between two levels. For example, the $(l_1,l_3)$ and $(l_3,l_1)$ edges
mean that $l_1$ and $l_3$ are at the same scale. Thus a spiritAgent can control several bodyAgents simultaneously present and activated in $l_1$ and $l_3$.
We note that $l_1\equiv l_3$

A \emph{loop} on a vertex indicates levels whose bodyAgents can adopt a similar behaviour. For example, a $(l_1,l_1)$ edge means that the spiritAgent, of some bodyAgents
situated in $l_1$, can be aggregated to form a single spiritAgent which will control these unchanged bodyAgents in $l_1$. These bodyAgents will have the same behaviour
when confronted to similar situations, but will keep their autonomy.

The following rules have to be applied if we want to obtain a coherent model.
\newtheorem{regle}{Rule}
\begin{regle}
Inclusion and Complementarity links are transitive.\\
$l_1 \prec l_2 \land l_2 \prec l_3 \to l_1 \prec l_3 $, $l_1 \equiv l_2 \land l_2 \equiv l_3 \to l_1 \equiv l_3 $.
\end{regle}
\begin{regle}
A level cannot be included in itself by a direct or transitive way.
This rule is translated by the fact that if we delete all pairs of symmetric edges, there should not be \emph{directed cycles} in the hierarchical level graph.\\
$\nexists l_1 \in L \land l_1 \prec l_1$
\end{regle}
\begin{regle}
Two distinct levels cannot share simultaneously an inclusion and a complementarity link, directly or by a transitive way.\\
$l_1\prec l_2 \to l_1\not\equiv l_2$, $l_1\equiv l_2\to l_1\not\prec l_2$.
\end{regle}

Each edge which is not part of a symmetric pair of edges is labelled with one or more aggregation function names.
An aggregation function name can be placed on several edges.

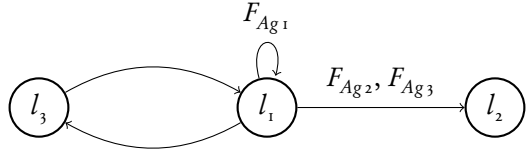
\begin{figure}[t]
\begin{center}

\begin{tikzpicture}

\node(l1)[shape=circle,thick,draw] at ( 0,0) {$l_1$};
\node(l2)[shape=circle,thick,draw] at ( 3,0) {$l_2$};
\node(l3)[shape=circle,thick,draw] at ( -3,0) {$l_3$};

\path	(l1)	edge [->, above]			node	{$F_{Ag2}$, $F_{Ag3}$}		(l2)
	     	edge [->, bend left]		node	{}				(l3)
	     	edge [loop above]  		node 	{$F_{Ag1}$}			(l1)
	(l3) 	edge [->, bend left]		node	{}				(l1);

\end{tikzpicture}

\end{center}
\caption{An example of Hierarchical Level Graph}
\end{figure}

The $(l_1,l_1)$ edge, labelled $F_{Ag1}$, indicates that the spiritAgents controlling some bodyAgents present in $l_1$ can aggregate themselves to form a single
spiritAgent controlling all these bodyAgents, through the $F_{Ag1}$ function.
The $(l_1,l_2)$ edge, labelled $F_{Ag2}$, $F_{Ag3}$, means that the spiritAgents controlling some bodyAgents present in $l_1$ can aggregate themselves to form a single
spriritAgent controlling a single aggregated bodyAgent situated in $l_2$, through the $F_{Ag2}$ or $F_{Ag3}$ function. 
These two functions concerns different combination of bodies.
And the symmetric pair of edges between $l_1$ and $l_3$, with no label, represents the fact that some spiritAgents can control simultaneously bodyAgents
situated in these two levels.

\subsection{Dis/Aggregation Functions}



\subsubsection{Content}

As shown before, there are two types of aggregation.
The first one deals with the aggregation of spiritAgents and the second one with the aggregation of spiritAgents and their associated bodyAgents.
The first type of aggregation is used to represent a set of agents with the same internal state,
that leads to agents which act similarly in the same situation but which can be place in several situations.
The aggregation of several bodyAgents without the aggregation of their spiritAgent is impossible because a body cannot be controlled
simultaneously by several concurrent spirits.

Once the hierarchical level graph is fixed, the modeler has to indicate every class of bodyAgent that he decides to place in 
levels and which class of spiritAgent control these bodyAgents.
For each aggregation function the modeler has to precise how many agents have to be merged, the class of 
aggregated and aggregate agents and how to generate internal and/or external state of the aggregate agent.

In this article we don't give any indication to set the decision module or the action module of aggregate agents or not but we focus on how to aggreagte
internal and external states of agents, respectively contained in spiritAgents and bodyAgents.
Each aggregation function can be divided into several subfunctions.
These subfunctions can be of two types. First type: a subfunction takes the same variable in each agents concerned
(spiritAgents or bodyAgents) and aggregates them to obtain a single value to place it in the aggregated agent state.
For example, a agent representing a platoon of vehicles has the mean position of all vehicle agents.
Second type: a subfunction similar to the first does an aggregation on several variables contained in the agents to aggregated
but produces only one value. This can be illustrated by the platoon agent described above. It only possesses one variable in its internal state called
``priority'' whose value is generated with the compound of the ``stamina'' and ``speed'' variables of each vehicle agents in the platoon.
Some variables of the agents to be aggregated can be ignored to construct an aggregate.\\

\subsubsection{Notation}

An aggregation function consists in creating a composite agent from several agents.
Here is the general form of an aggregation function $F_{Ag}$ using for argument $n$ conceptualAgent class, $cta$ (class to aggregate),
endowed of an interval, $[min_i, max_i]$, indicating how many instances of these classes are necessary to accomplish this aggregation.
For each conceptualAgent class it is precised if the aggregation implies bodyAgents in addition of spiritAgent
with the indication of a level $l_i$ where the bodyAgents are situated.
The class of the agent produced by the aggregation, $AAC$ (Aggregate Agent Class), is the output of $F_{Ag}$ with its level $l$ if the aggregation concerns bodyAgents.
If the aggregation only concerns spiritAgents $l = l_i = \varnothing$.\\

\begin{equation}
F_{Ag} ( \prod_{i \in n} \langle[min_{i};max_{i}]cta_i, l_i\rangle) = (AAC, l)\\
\end{equation}

For example, let consider the $F_{Ag2}$ function described in the hierarchical graph below.
Let $F_{Ag2}$ aggregates one bodyAgent of class \emph{Leader} and at least 4 to 9 
bodyAgents of class \emph{Follower} all situated in $l_1$ level and their linked spiritAgents to create
a bodyAgent of class \emph{Platoon} situated in $l_2$ level and its linked spiritAgent. Then:\\

\begin{equation}
F_{Ag2} (\langle[1;1], Leader, l_1\rangle, \langle[4;9], Follower, l_1\rangle) = (Platoon, l_2)\\
\end{equation}

Aggregation subfunctions have quite the same notation than aggregation functions.
It is not necessary to precise the number of concerned agents anymore. But variables,
in concerned agents, which will be mixed together have to be known.
For example the subfunction described in the previous subsection can be noted like this:

\begin{equation}
\begin{array}{rcl}
f_{Ag2,1} \big((Leader.stamina, Leader.speed, l_1),\\
(Follower.stamina, Follower.speed, l_1)\big)\\
 = (Crowd.priority, l_2)\\
\end{array}
\end{equation}

\subsubsection{Disaggregation and Memorization Functions}

Each aggregation function possesses its disaggregation function and eventually a memorization function.
A disaggregation function permits to create several instances of the aggregated agents from the aggregate agent.
A memorization function can be used to store some information.
Each memorization function is associated to a disaggregation one to generate several agents representing the initial aggregated agents
taking into account the last state of the aggregated agents and the system evolution since the aggregation.
Here, $nb_i$ indicates the number of agents of each class involved in the aggregation.

\begin{equation}
\begin{array}{rcl}
F_{Disag}(AAC, l, F_{Memorization}(\prod_{i \in n} \langle nb_i,cta_i, l_i\rangle) )
=( \prod_{i \in n} \langle nb_i,cta_i, l_i\rangle)
\end{array}
\end{equation}

These two functions are divided in subfunctions in a similar way than the aggregation function.
Let take a platoon endowed of the two position variables, $X$ and $Y$, representing the position variable $x$ and $y$ of all the vehicles constituting it.
The memorization function store positions of all these vehicles. Memorization is not active during the execution of the platoon agent.
After the platoon agent have moved in $(X',Y')$ position, it can be disaggregated by recreating the vehicles agents, calculating the value of their $x$ and $y$ variables
with $X'$ and $Y'$ and applying the memorized repartition.

\subsection{Dis/Aggregation Tests}

\citet{Navarro:2011} explains how to decide when agents should be aggregated.
He uses an affinity function which measure the similarity of internal and external states of agents.
When the similarity is more important than a given threshold he links the two agents.
Linked agents with the higher similarity value are aggregated together.

We can use a similar mechanism to decide when to use an aggregation function, but in our case we need one utility function $Aff$ by aggregation function $F_{Ag}$.
If there are several aggregation functions which concern the same spiritAgents or bodyAgents in the same levels,
it is necessary to decide when apply one instead of another. There are three possibilities.
1) The choice of $F_{Ag}$ is done after measuring the affinity of agent groups with all $Aff$
and the aggregate are instantiated each time, choosing the group with the higher affinity, until there is no group.
2) It is also possible to impose an order to test different $F_{Ag}$. All groups with a high affinity for one  $F_{Ag}$
are aggregated, then the next  $F_{Ag}$ is tested until there is no more  $F_{Ag}$.
3) The choice of $F_{Ag}$ can be done by a mix of the two previous methods. An partial order is defined on $F_{Ag}s$ space.
And if there is no precedence link between different $F_{Ag}$, we apply the first method to aggregate agents considering that the model $F_{Ag}$
only contains these $F_{Ag}$ after that we continue following the established order.

\section{Measuring Consistency}

\citet{Davis:1993} uses the notion of consistency to measure the quality of simulations dealing with models of different resolution.
``Consistency between a high-resolution model $\textbf{M}$ and a low-resolution model $\textbf{M'}$
is the comparison between the projected state of an aggregate of high-resolution entities which evolved in $\textbf{M}$, and
the projected state of the same aggregate initially controlled by $\textbf{M'}$\,''.

It is more intuitive to base the comparison on the evolution of the more detailed model instead of the aggregate model
because it has a higher resolution and possesses more significant information.

\begin{figure}[t]
\begin{center}

\begin{tikzpicture}
\node(a) at (0,0){
            \begin{tabular}{c}
	High-resolution \\
	inputs
\end{tabular}};
\node[draw] at (4,0)(b){
            \begin{tabular}{c}
	High-resolution \\
	model : \textbf{M}
\end{tabular}}
edge [-,ultra thick] (a);
\node(c) at (8,0){
            \begin{tabular}{c}
	High-resolution \\
	outputs
\end{tabular}}
edge [<-,ultra thick] (b);

\draw (0,-2) node[draw](d){
            \begin{tabular}{c}
	Aggregate \\
	function
\end{tabular}}
edge [-,ultra thick] (a);
\draw (8,-2) node[draw](e){
            \begin{tabular}{c}
	Aggregate \\
	function
\end{tabular}}
edge [-,ultra thick] (c);

\draw (8,-4.7) node(f){
            \begin{tabular}{c}
	Aggregated \\
	high-resolution\\
	outputs
\end{tabular}}
edge [<-,ultra thick] (e);

\draw (0,-6) node(g){
            \begin{tabular}{c}
	Low-resolution \\
	inputs
\end{tabular}}
edge [<-,ultra thick] (d);
\draw (4,-6) node[draw](h){
            \begin{tabular}{c}
	Low-resolution \\
	model : \textbf{M'}
\end{tabular}}
edge [-,ultra thick] (g);
\draw (8,-6) node(i){
            \begin{tabular}{c}
	Low-resolution \\
	outputs
\end{tabular}}
edge [<-,ultra thick] (h);

\draw[ thick,decorate,decoration=brace] (9.2,-4) -- (9.2,-6.5);

\draw (11,-5.3) node(j){
            \begin{tabular}{l}
	The models are \\
	consistent if these\\
	are approximately\\
	equal
\end{tabular}};

\end{tikzpicture}

\end{center}
\caption{Weak consistency, according to~\citet{Davis:1993}}
\end{figure}
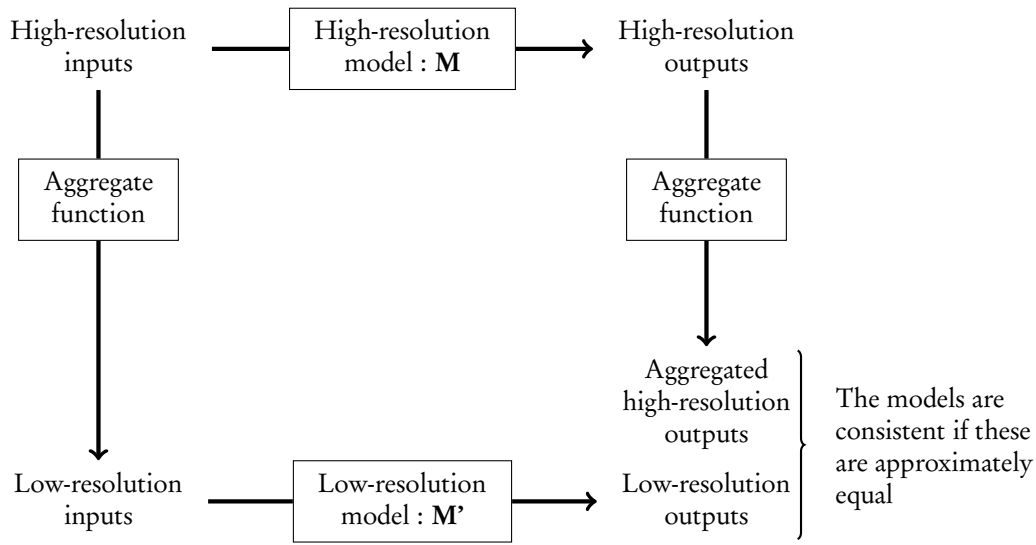


Before modeling the system, it is necessary to locate the significant simulation elements.
These elements can be in the internal (spiritAgent) or external (bodyAgent) states of agents or in their environment.
Once these elements are identified, several simulations are launched with the same parameters (initial state and execution time)
using only the most detailed levels, carrying the more information but the most expensive one.
At the end of the simulations execution a mean state of the identified elements is recorded.
The same process is done with the model using dynamic change of LOD. Then the dissimilarity is measure
between these two recording to calculate the consistency.

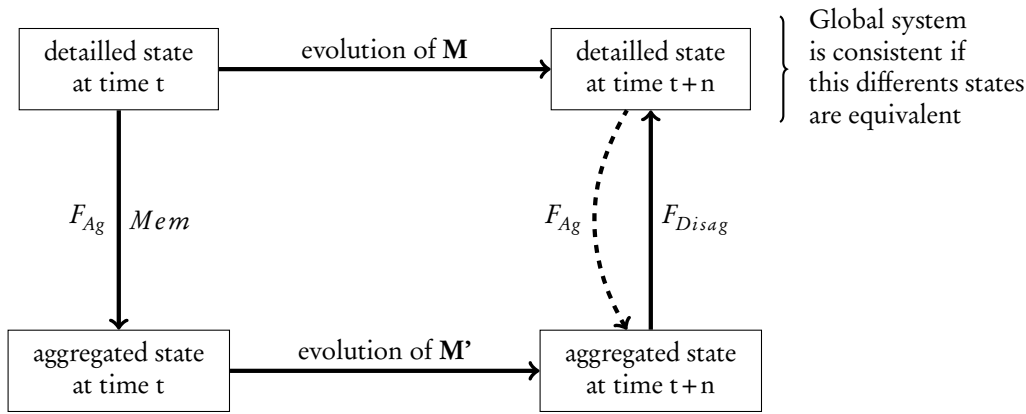
\begin{figure}[t]
\begin{center}

\begin{tikzpicture}
\node(a)[draw] at (0,0){
            \begin{tabular}{c}
	detailled state\\
	at time t
\end{tabular}};
edge [-,ultra thick] (a);
\node(b)[draw] at (7,0){
            \begin{tabular}{c}
	detailled state\\
	at time t+n
\end{tabular}}
edge [<-,ultra thick]node[swap, above]{evolution of \textbf{M}} (a);

\draw (0,-4)node(c)[draw] {
            \begin{tabular}{c}
	aggregated state\\
	at time t
\end{tabular}}
edge [<-,ultra thick]node[swap,left]{$F_{Ag}$}node[swap,right]{$Mem$}   (a);
\draw (7,-4) node(d)[draw]{
            \begin{tabular}{c}
	aggregated state\\
	at time t+n
\end{tabular}}
edge [<-,ultra thick]node[swap, above]{evolution of \textbf{M'}}  (c)
edge [->,ultra thick] node[swap,right]{$F_{Disag}$}(b)
edge [<-, dashed,ultra thick, bend left]node[swap,left]{$F_{Ag}$} (b)
;

\draw[ thick,decorate,decoration=brace] (8.7,0.7) -- (8.7,-0.7);

\draw (10.5,0) node(j){
            \begin{tabular}{l}
	Global system \\
	is consistent if\\
	this differents states\\
	are equivalent
\end{tabular}};

\end{tikzpicture}

\end{center}
\caption{Strong consistency, according to~\citet{Davis:1993}}
\end{figure}

\section{Conclusion and Perspectives}

This article introduces a methodology and theoretical tools to engineer and validate multi-level agent based simulations with dynamic change of LOD.

It is applied in the european project InTrade\footnote{http://www.intrade-nwe.eu/}.
This project deals with logistic in european container ports endowed with Autonomous Intelligent Vehicles (AIV).
Partners involved in this project work at different scales and use simulation tools adapted to it (SCANeRstudio or Flexsim Container Terminal
\footnote{http://www.intrade-nwe.eu/ or www.flexsim.com/}).
The agent-based platform MadKit\footnote{http://www.madkit.org/} is used to make models coexist in a single simulation.
Results are visualized with SCANeRstudio or Flexsim CT.

An interesting perspective of this work would be to find better ways (cheaper or more realistic) to decide when simulated entities should be (dis)aggregated. It is closely related to the emergence detection and reification problem~\citep{David:2009}. Two main approaches have been proposed to tackle this issue: a statistical one (\textit{e.g.},~\citep{Caillou:2012,Caillou:2013,Moncion:2010,Vo:2012}) and a symbolic one~\citep{Chen:2009,Chen:2010}. It would be interesting to integrate them.

Another perspective is the integration of organizational concepts, such as \textit{Systems of Systems} (SoS), in our methodology. It would allow to explicitly represent system or group level properties such as goals or missions.


\bibliographystyle{apalike}
{\footnotesize \bibliography{maBiblio}}

\end{document}